\documentclass[12pt]{article}

\usepackage[english]{babel}

\usepackage[letterpaper,top=2cm,bottom=2cm,left=3cm,right=3cm,marginparwidth=1.75cm]{geometry}

\usepackage{amsmath}
\usepackage{graphicx,natbib}
\usepackage[colorlinks=true, allcolors=blue]{hyperref}
\usepackage{multirow}
\usepackage{graphics,booktabs}
\usepackage{amsthm,amssymb,authblk}
\usepackage{relsize}

\theoremstyle{definition}

\newtheorem{remark}{Remark}[section]

\usepackage{algorithm,algpseudocode}

\title{Optimal subdata selection for linear model selection}
\author{Vasilis Chasiotis}
\author{Dimitris Karlis}

\affil{\small Department of Statistics, Athens University of Economics and Business,  Greece}

\date{}

\usepackage{lipsum}

\begin{document}
	\maketitle
	
	\begin{abstract}
		If the assumed model does not accurately capture the underlying structure of the data, a statistical method is likely to yield sub-optimal results, and so model selection is crucial in order to conduct any statistical analysis. However, in case of massive datasets, the selection of an appropriate model from a large pool of candidates becomes computationally challenging, and limited research has been conducted on data selection for model selection. In this study, we conduct subdata selection based on the A-optimality criterion, allowing to perform model selection on a smaller subset of the data. We evaluate our approach based on the probability of selecting the best model and on the estimation efficiency through simulation experiments and two real data applications.
	\end{abstract}
	
	
	\section{Introduction}
	\label{introduction}
	The field of research across various disciplines is currently facing an unprecedented demand for analyzing large-scale datasets, commonly referred to as big data. As the size of these datasets continues to grow, so does the need for substantial computational resources to facilitate the statistical analysis and modeling processes. Despite the rapid advancements in computational power, the surge in data volume has outpaced the availability of resources.
	
	Consequently, this presents new and significant challenges in terms of data storage and analysis. To tackle these challenges, a common strategy involves data reduction or subsampling techniques, where a subset of the data is selected to extract relevant information. This step plays a pivotal role in the analysis of big data. By employing subsampling techniques, the computational burden is alleviated by significantly reducing the dataset size, thus making it more manageable and feasible for analysis.
	
	In recent years, there is an increasing interest in subdata selection through a deterministic way. At first, \cite{wang2019information}, interested in estimating the unknown parameters of a linear regression model, proposed the information-based optimal subdata selection, which is based on D-optimal designs. Also, for further related work, based on the concept of optimal experimental designs, were focused on estimating the unknown parameters, see \cite{wang2019divide}; \cite{deldossi2021optimal}; \cite{wang2021oss}; \cite{chasiotis2023}; \cite{chasiotis2023levss}. More information on subdata selection or subsampling from big data based on designs can be found in the review papers by \cite{yao2021review} and \cite{yu2023}, which provides a comprehensive overview of the current state of research in this area. Also, 
	
	It is well known that model selection is particularly important in case of regression problems. The work of \cite{yu2022} provides an overview about different methods related to the model selection problem. However, when dealing with massive datasets, significant challenges can be posed, particularly when the full data size is large, and the number of candidate models is moderately large. When uncertainty exists regarding the choice of the model, limited research has been conducted on how to effectively perform data selection under such circumstances. \cite{yu2022} provided  a deterministic selection of the most informative data points from the full data based on leverages scores (LEVSS) for linear model discrimination. Also, \cite{singh} introduced a method that combines Lasso regression and subdata selection aiming at both variable selection and building a predictive model, when the full data size is very large, and the number of variables is large. Moreover, we should refer to the work of \cite{deldossi} who proposed informative and non-informative subdata selection methods, that is if information about the responses is available or not, respectively. Also, they highlighted the importance of the I-optimal designs when the goal is to get accurate predictions on a specified prediction set.
	
	In the current paper, we are interested in selecting subdata that allows us to perform model selection. We focus on the selection of the subdata based on the A-optimality criterion that is widely used in the area of optimal experimental designs. Our motivation is the work of \cite{jones} who proposed the choice of A-optimal designs for screening experiments. The remaining of the paper proceeds as follows. Section \ref{theor} provides some notations as well as the BIC criterion. Section \ref{meth} describes our motivation, and two proposed algorithms are provided as well. Simulation evidence to support our proposed approach is provided in Section \ref{section_simulation}. A comparison with LEVSS approach is included, in order to show the improvement gained. Two real datasets are used in Section \ref{section_real_data}, while concluding remarks can be found in Section \ref{concluding_remarks}.

	\section{Framework}\label{theor}
	Let $y_1, \ldots, y_n$ be independent responses that have been generated by
	\begin{equation}\label{model1}
		y_i = \mu_i + \varepsilon_i, \quad i = 1, 2, \ldots, n,    
	\end{equation}
	where $\mu_i = \textbf{x}_i^\text{T} \boldsymbol{\beta}$, $\textbf{x}_i=(x_{i1},x_{i2},\ldots,x_{ip})^{\text{T}}$ is a covariate vector, $\boldsymbol{\beta}=(\beta_1,\beta_2,\ldots,\beta_p)^{\text{T}}$ is a $p$-dimensional vector of unknown slope parameters and $\varepsilon_i$ is an error term such that E$(\varepsilon_i)=0$ and V$(\varepsilon_i)=\sigma^2$. Also, the $y_i$'s are uncorrelated given covariates $\textbf{x}_i$, $i=1,2,\ldots,n$.
	
	We assume that the true model for $\mu_i$ lies in a candidate set of models. To facilitate the presentation, let $\textbf{y} = (y_1, \ldots, y_n)^\text{T}$, $\boldsymbol{\mu} = (\mu_1, \ldots, \mu_n)^\text{T}$, and  $\textbf{X} = (x_1^\text{T}, \ldots, x_n^\text{T})^\text{T}$. 
	
	In model \eqref{model1}, the intercept parameter, denoted as $\beta_0$, has been omitted, since in model selection problems, it is often not of interest. The elimination of $\beta_0$ in linear regression models can be achieved by centralizing the full data $(\textbf{y}, \textbf{X})$, and so $\boldsymbol{\beta}$ can be estimated by fitting the least squares on the centralized full data without $\beta_0$. However, in case of prediction, $\beta_0$ is relevant, and so it can be estimated with the adjusted estimator $\hat{\beta}_0=\bar{\textbf{y}}-\bar{\textbf{X}}^{\text{T}}\hat{\boldsymbol{\beta}}$ \citep{wang2019information}, where $\bar{\textbf{y}}$ is the mean of $\textbf{y}$, $\bar{\textbf{X}}$ is the column mean vector of $\textbf{X}$, and $\hat{\beta}$ is the least squares estimate of $\boldsymbol{\beta}$ from the centralized data. Note that \cite{wang2022center} showed that the least squares estimator on $\boldsymbol{\beta}$ obtained from a model without $\beta_0$ is unbiased and it has a smaller variance covariance matrix in the Loewner order than that obtained from a model with $\beta_0$.
	
	Let $M_r$, $r = 1, 2, \ldots, 2^p$ be a subset of the column indices of $\textbf{X}$, that is the number of candidate models that we take into consideration is equal to $2^p$. Therefore, the candidate model corresponding to $M_r$ has the form
	\begin{equation}\label{mu}
		\boldsymbol{\mu} = \textbf{X}^{(r)} \beta^{(r)},
	\end{equation}
	where $\textbf{X}^{(r)}$ is an $n \times p^{(r)}$ submatrix of $\textbf{X}$ and $\boldsymbol{\beta}^{(r)}$ is the vector of unknown slope parameters. 
	
	We use $\mathcal{S}$ to denote the set of all candidate models, assuming that the $2^p$-th model is the widest one that it is always included in $\mathcal{S}$. The $r$-th candidate model is said to be correct if $\boldsymbol{\beta}^{(r)}$ satisfies that $\boldsymbol{\mu} = \textbf{X}^{(r)}\boldsymbol{\beta}^{(r)} 
	$. Let $\mathcal{S}_c$  be the set of correct candidate models. In case that the number of models in $\mathcal{S}_c$ is more than two, then the selected model is the one with the fewest parameters, also called as the true model.
	
	For model \eqref{mu}, the $r$-th candidate model according to the Bayesian information criterion (BIC) based on the full data $(\textbf{y}, \textbf{X})$ is written as
	\begin{equation}\label{bic1}
		\hat{M}_{\text{BIC}} = \arg\min_{M_r} \left\{ n \log\left(\frac{1}{n}\sum_{i=1}^n \left(y_i - \hat{\mu}_i^{(r)}\right)^2\right) + p^{(r)} \log(n) \right\},    
	\end{equation}
	where $p^{(r)}$ is the cardinality of $M_r$, $\hat{\mu}_i^{(r)}$ is the $i$-th element of $\hat{\boldsymbol{\mu}}^{(r)} = \textbf{X}^{(r)} \hat{\boldsymbol{\beta}}^{(r)}$, and $\hat{\boldsymbol{\beta}}^{(r)}$ is the maximum likelihood estimator under the $r$-th candidate model.
	
	Under the selection of subdata of size $k$, the selected model according to BIC in \eqref{bic1}, denoted as $\hat{M}^*_{\text{BIC}}$, satisfies that
	\begin{equation}\label{bic2}
		\hat{M}^*_{\text{BIC}} = \arg\min_{M_r}\left\{k \log\left(\frac{1}{k}\sum_{i=1}^k\left(y^*_i - \hat{\mu}_i^{*(r)}\right)^2\right) + p^{(r)} \log(k) \right\}    
	\end{equation}
	where $\hat{\mu}_i^{*(r)}$ is the $i$-th element of $\hat{\boldsymbol{\mu}}^{*(r)} = \textbf{X}^*_{(r)} \hat{\boldsymbol{\beta}}^*_{(r)}$, $\textbf{X}^*_{(r)}$ is the $k \times p^{(r)}$ submatrix of $\textbf{X}$ for the selected subdata, and $\hat{\boldsymbol{\beta}}^*_{(r)}$ is the maximum likelihood estimator of the $r$-th candidate model under the selected subdata.
	
	\section{Methodology}\label{meth}
	We are interested in both model selection and as much as possible good estimation results based on the selected model under the subdata selection. Also, we assume that any information about the responses is not available or is too expensive to have it.
	
	\cite{yu2022} proved that both selecting a good model and achieving better estimation results
	using the selected model based on the subdata, can be achieved by selecting subdata for which the  determinant of the corresponding information matrix is maximized, that is one should select subdata that are D-optimal. However, we are motivated by the work of \cite{jones}, who proposed the choice of A-optimal designs rather than D-optimal designs for screening experiments. The goal of a screening experiment is the identification of the experimental variables that have a real influence on the result, that is to identify an active subset of the factors. It can be seen that a relation between screening designs and model selection exists. A design that minimizes the average variance of the parameter estimates is called A-optimal. This can be achieved by minimizing the trace of the inverse of the information matrix. Thus, subdata that minimizes the trace of the inverse of the corresponding information matrix over all possible selected subdata are called A-optimal.
	
	The challenge is to find a computationally fast algorithm to obtain subdata that are as close as possible to the exact A-optimal ones. It is obviously infeasible to obtain exact A-optimal subdata, since an exhaustive algorithm will be time-consuming in case of big data. Therefore, we are interested in eliminating a proportion of the full data and selecting the A-optimal subdata among the remaining ones. Our primary tool to achieve our goal is the Sherman-Morrison formula:
	\begin{equation}\label{smf}
		\left(\textbf{X}^{-\text{T}}_i \textbf{X}^{-}_i\right)^{-1} = \left(\textbf{X}^\text{T} \textbf{X}\right)^{-1} + \dfrac{\left\lbrace\left(\textbf{X}^\text{T} \textbf{X}\right)^{-1}\textbf{x}_i\right\rbrace\left\lbrace\left(\textbf{X}^\text{T} \textbf{X}\right)^{-1}\textbf{x}_i\right\rbrace^\text{T}}{1-\textbf{x}_i^\text{T}\left(\textbf{X}^\text{T} \textbf{X}\right)^{-1} \textbf{x}_i},
	\end{equation}
	where $\textbf{X}^{-}_i$ is the design matrix attained by removing the $i$-th row, that is the $p\times1$ covariate vector $\textbf{x}_i$, from $\textbf{X}$. Note that the denominator of the fraction in the right side of \eqref{smf} cam be written as $1-h_{ii}$, where $h_{ii}$ is the $i$-th diagonal element of the hat matrix of $\textbf{X}$.
	
	Therefore, we get that
	\begin{equation*}
		\text{tr}\left\lbrace\left(\textbf{X}^{-\text{T}}_i \textbf{X}^{-}_i\right)^{-1}\right\rbrace= \text{tr}\left\lbrace\left(\textbf{X}^\text{T} \textbf{X}\right)^{-1}\right\rbrace + \dfrac{\text{tr}\left[\left\lbrace\left(\textbf{X}^\text{T} \textbf{X}\right)^{-1}\textbf{x}_i\right\rbrace\left\lbrace\left(\textbf{X}^\text{T} \textbf{X}\right)^{-1}\textbf{x}_i\right\rbrace^\text{T}\right]}{1-h_{ii}}
	\end{equation*}
	or
	\begin{equation}\label{smftr}
		\text{tr}\left\lbrace\left(\textbf{X}^{-\text{T}}_i \textbf{X}^{-}_i\right)^{-1}\right\rbrace= \text{tr}\left\lbrace\left(\textbf{X}^\text{T} \textbf{X}\right)^{-1}\right\rbrace + \dfrac{\mathlarger{\sum}_{j=1}^{p}z_j^2}{1-h_{ii}},
	\end{equation}
	where tr($\cdot$) denotes the trace of a square matrix, and $z_j$ is the $j$-th element of the $p\times1$ vector $\left(\textbf{X}^\text{T} \textbf{X}\right)^{-1}\textbf{x}_i$.
	
	As it was expected, we get according to \eqref{smftr} that  $\text{tr}\left\lbrace\left(\textbf{X}^{-\text{T}}_i \textbf{X}^{-}_i\right)^{-1}\right\rbrace>\text{tr}\left\lbrace\left(\textbf{X}^\text{T} \textbf{X}\right)^{-1}\right\rbrace$. Therefore, we should delete data points from the full data, trying the value of \linebreak $\text{tr}\left\lbrace\left(\textbf{X}^{-\text{T}}_i \textbf{X}^{-}_i\right)^{-1}\right\rbrace$ to be as close as possible to the value of $\text{tr}\left\lbrace\left(\textbf{X}^\text{T} \textbf{X}\right)^{-1}\right\rbrace$. One could obtain the final subdata, deleting data points one after another from the full data, based on \eqref{smftr}, but in such case the algorithm will be time consuming. Thus, we should firstly benefit from \eqref{smftr} to conduct a generous elimination in the full data. Then, we could apply \eqref{smftr} in the remaining data points, conducting a second more precise elimination procedure, in order to lead to the selection of the A-optimal subdata.
	
	According to \eqref{smftr}, one should take into consideration for removing the covariate vectors $\textbf{x}_i$'s from $\textbf{X}$ for which the value of the corresponding $h_{ii}$ are small. This is the fact in which we rely on for the first proposed algorithms. Also, it is interesting to take into consideration the value of $\mathlarger{\sum}_{j=1}^{p}z_j^2$ in \eqref{smftr}, examining its effect on $h_{ii}$'s. Therefore, the elimination step for the second proposed algorithm is based on removing the covariate vectors $\textbf{x}_i$'s from $\textbf{X}$ for which the value of the fraction in the right side of \eqref{smf} is small. A precise consideration of \eqref{smftr} takes place in the main parts of both algorithms.
	
	\begin{algorithm}
		\caption{}
		\begin{algorithmic}
			\Require The design matrix $\textbf{X}=(\textbf{x}_{i}^\text{T}), i=1,2,\ldots,n$ and the target sample size ($k>p$).
			\Ensure The selected subdata $\textbf{Q}$.
			\State \textbf{Step 1: Elimination}
			\State $\textbf{Q}$ 
			\Comment{$2k\times p$ design matrix obtained by LEVSS algorithm}
			\State \textbf{Step 2: Preparation}
			\State $\text{INF}= \left(\textbf{Q}^{\text{T}}\textbf{Q}\right)^{-1}$
			\State $\text{tINF}= \text{tr}\left\lbrace\left(\textbf{Q}^{\text{T}}\textbf{Q}\right)^{-1}\right\rbrace$
			\State $N = \text{nrow} \left(\textbf{Q}\right)$
			\Comment{number of data points in $\textbf{Q}$}
			\State \textbf{Step 3: Main algorithm}
			\For{$j$ in $1,2,\ldots,k$}
			\For{$i$ in $1,2,\ldots,N$}
			\State $\textbf{S}_i=\left(\textbf{Q}_i^{-\text{T}}\textbf{Q}_i^{-}\right)^{-1}$ \Comment{compute for $\textbf{x}_i$ based on \eqref{smf} considering INF}
			\State $s_i=\text{tr}\left\lbrace\left(\textbf{Q}_i^{-\text{T}}\textbf{Q}_i^{-}\right)^{-1}\right\rbrace$ \Comment{compute for $\textbf{x}_i$ based on \eqref{smftr} considering tINF}
			\EndFor
			\State tINF = min($s_i$)
			\State INF = $\left(\textbf{Q}_i^{-\text{T}}\textbf{Q}_i^{-}\right)^{-1}$
			\Comment{keep $\left(\textbf{Q}_i^{-\text{T}}\textbf{Q}_i^{-}\right)^{-1}$ whose $s_i$ is the minimum}
			\State $\textbf{Q}=\textbf{Q}^{-}_{i}$
			\Comment{delete $\textbf{x}_{i}^\text{T}$ from \textbf{Q} whose $s_i$ is the minimum}
			\State $N = N -1$
			\EndFor
		\end{algorithmic}
	\end{algorithm}
	
	\begin{algorithm}
		\caption{}
		\begin{algorithmic}
			\Require The design matrix $\textbf{X}=(\textbf{x}_{i}^\text{T}), i=1,2,\ldots,n$ and the target sample size ($k>p$).
			\Ensure The selected subdata $\textbf{Q}$.
			\State \textbf{Step 1: Elimination}
			\State $\textbf{X}=\textbf{UD}\textbf{V}^\text{T}$ 
			\Comment{perform singular value decomposition of \textbf{X}}
			\State $h_{ii}=\lVert U_{i\cdot}\rVert^2$
			\Comment{$U_{i\cdot}$ denotes the $i$-th row of $\mathbf{U}$}
			\State $d_i=\mathlarger{\sum}_{j=1}^{p}z_j^2/(1-h_{ii})$
			\Comment{ $z_j$ is the $j$-th element of $\textbf{V}\textbf{D}^2\textbf{V}^\text{T}\textbf{x}_{i}$}
			\State \textbf{Q}
			\Comment{$2k\times p$ design matrix keeping $\textbf{x}_{i}^\text{T}$'s whose $d_i$ are the maximum}
			\State \textbf{Steps 2 and 3: Same as in Algorithm 1}
		\end{algorithmic}
	\end{algorithm}
	
	\begin{remark}\label{rem1}
		As \cite{jones} stated, in screening designs, it is a common practice to scale each quantitative factor so that its minimum value corresponds to $-1$ and its maximum value corresponds to $1$. By doing so, all the quantitative factors are standardized and brought to a common scale. This scaling ensures that the factors are comparable and eliminates any potential bias that may arise from differences in the original scales of the factors. Therefore, in Steps 2 and 3 of Algorithm 1 and in all steps of Algorithm 2, the data are scaled to $[-1,1]$.
	\end{remark}
	
	\begin{remark}\label{rem2}
		From the $j$-th to the ($j+1$)-th iteration, the new obtain design matrix \textbf{Q} is one among the $\textbf{Q}^-_{i}$'s, $i=1,2,\ldots,N$, which is used in \eqref{smf} and \eqref{smftr} as the design matrix \textbf{X}.
	\end{remark}
	
	\begin{remark}\label{rem3}
		Algorithm 1 is based on the algorithm of the LEVSS approach (see the elimination step). Therefore, in case the algorithm of the LEVSS approach selects more than $k$ data points, say $k^*$, then in the elimination step of Algorithm 1, $2k$ out of $k^*$ data points are selected with a simple random sampling. 
	\end{remark}
	
	\begin{remark}\label{rem4}
		Since in both algorithms we firstly keep $2k$ data points, interested in selecting subdata whose size is equal to $k$, in Step 3 for both, we further delete $k$ data points. Of course, one could keep more or less than $2k$ data points, but the information gain and/or the computational cost should be considered.
	\end{remark}
	
	One can easily see the importance of Sherman-Morrison formula, because the inverse of the information matrix is computed only at the beginning of the preparation step in both algorithms. In the main part of both algorithms, in the ($j+1$)-th iteration, the inverse of the information matrix computed in the $j$-th iteration is used.
	
\begin{figure}[!thb]
		\centering
		\includegraphics[width=1\textwidth]{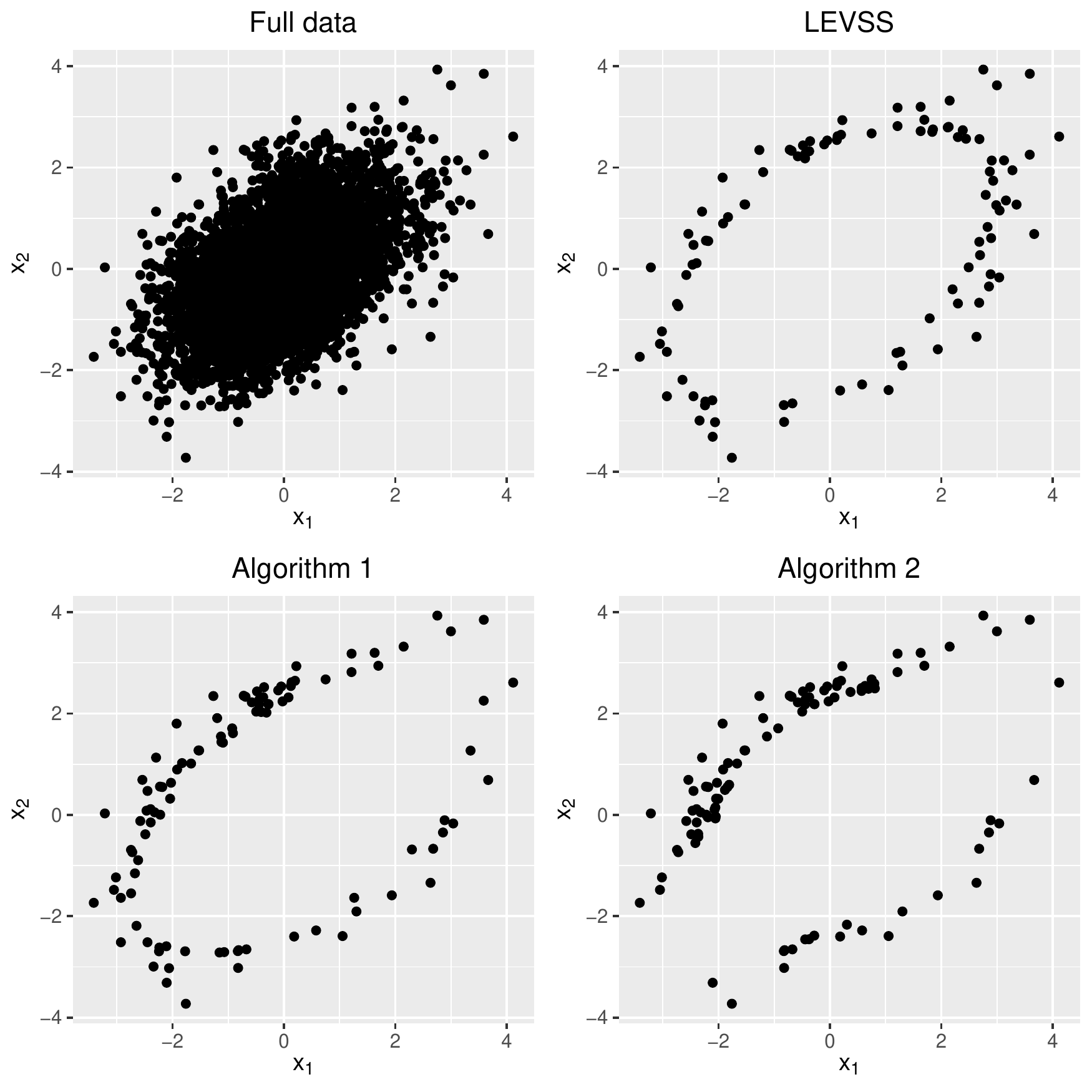}
		\caption{A toy example for the different approaches. Data with two covariates and full data size of $5000$ data points were generated. The different approaches were used to select $100$ data points. In the first row, the full data can be seen in the first panel and the LEVSS approach in the second one. In the second row, Algorithms 1 and 2 of our proposed approach can be seen in the first and the second panel, respectively.}
		\label{toy}
	\end{figure}

	In Figure \ref{toy} we present the selected subdata according to Algorithms 1 and 2 as well as the algorithm of the LEVSS approach in a toy example. The full data are presented as well. We have used two covariates and the size of the full data is $5000$. Observations $\textbf{x}_1$ and $\textbf{x}_2$ follow a multivariate normal distribution, that is, $\textbf{x}_i\sim N(\textbf{0},\mathbf{\Sigma})$, where $\mathbf{\Sigma}=\left(\Sigma_{ij}\right)$, $i,j=1,2$ is a covariance matrix. Also, $\Sigma_{ij}=1$ for $i=j=1,2$ and $\Sigma_{ij}=0.5$ for $i\ne j=1,2$. Suppose that we are interested in selecting $100$ observations. We can see that Algorithms 1 and 2
	select data points that are very similar to the ones selected by the algorithm of the LEVSS approach. However, the subdata selected by the algorithm of the LEVSS approach are more ``uniformly distributed'' in the convex hull of the full data compared to the ones selected by Algorithms 1 and 2.
	
	\section{Simulation experiments}\label{section_simulation}
	In this section, we evaluate the performance of Algorithms 1 and 2 based on simulated data, presenting the results of  the LEVSS approach as well, to make a comparison.
	
	We use the same setting as in \cite{yu2022}, that is there are seven covariates to be considered, and an intercept term with the true value fixed at $\beta_{0} = 0.25$. Also, we generate 
	$$\mu_i = \beta_{0} + \sum_{j=1}^{p}x_{ij}\beta_j, i = 1, \ldots, n.$$ 
	The responses $y_i$'s are generated from 
	$$y_i = \mu_i + \varepsilon_i$$ with $\sigma^2=1$. For the slope parameters in each repetition of the simulation, the values of $\beta_1$ and $\beta_2$ are independently generated from the distribution $\text{Unif}(0.5, 1)$, the values of $\beta_3$ and $\beta_4$ are independently generated from the distribution $\text{Unif}(0.05, 0.1)$, and we set $\beta_j = 0$ for $j = 5, 6, 7$. Also, assuming that any non-empty subset of these variables can be a candidate set of active variables, there are $2^{7} - 1 = 127$ candidate models. 
	
	Our method is illustrated using all-subset regression. Since the results via forward regression are quite similar, we provide them in the Appendix. 
	
	The covariates $\textbf{x}_i = (x_{i1}, x_{i2}, \ldots, x_{i7})^{T}$ for the full data with $n = 500,000$ are generated from:
	
	\begin{enumerate}
		\item[] Case 1: Multivariate normal distribution $N(0,\boldsymbol{\Sigma}_1)$ with the $(i,j)$-th entry of $\boldsymbol{\Sigma}_1$ being $0.5^{|i-j|}$.
		
		\item[] Case 2: Multivariate normal distribution $N(0,\boldsymbol{\Sigma}_2)$ with the $(i,j)$-th entry of $\boldsymbol{\Sigma}_2$ being $0.5^{I(i \neq j)}$, where $I(\cdot)$ is the indicator function.
		
		\item[] Case 3: Mixture multivariate normal distribution $0.5N(0,\boldsymbol{\Sigma}_1) + 0.5N(0,\boldsymbol{\Sigma}_2)$, where $\boldsymbol{\Sigma}_1$ and $\boldsymbol{\Sigma}_2$ are defined in Cases 1 and 2, respectively.
		
		\item[] Case 4: Multivariate $t$-distribution with three degrees of freedom. The mean parameter is $0$, and the scale matrix parameter $\boldsymbol{\Sigma}_1$ is defined in Case 1.
		
		\item[] Case 5: Multivariate $t$-distribution with three degrees of freedom. The mean parameter is $0$, and the scale matrix parameter $\boldsymbol{\Sigma}_2$ is defined in Case 2.
		
		\item[] Case 6: Log-normal distribution with parameters $0$ and $\boldsymbol{\Sigma}_2$, which is defined in Case 2. 
	\end{enumerate}
	
	For each candidate model $M_r$, the slope parameters $\hat{\beta}^*_{(r)}$ are estimated using the selected subdata from the centralized full data, and then the intercept term $\beta_{0(r)}$ is estimated by $\bar{\textbf{y}} - \bar{\textbf{X}}^T_{(k)}\hat{\beta}^*_{(r)}$.
	
	We evaluate our proposed subdata selection approach by the following two criteria:
	\begin{enumerate}
		\item Accuracy, that is the probability of selecting the true model, via BIC (see \eqref{bic2})
		\item The mean squared prediction error (MSPE) for the observations in the test sample, that is 
		$$\text{MSPE} = \dfrac{1}{n_{\text{t}}}\sum_{i=1}^{n_{\text{t}}} \lVert\mu_{i,\text{t}} - \hat{\beta}_{0(r)}^* - \textbf{x}_{i(r),\text{t}}^T \hat{\beta}^*_{(r)}\rVert^2,$$ where $n_{\text{t}}$ is the size of the test data, $\mu_{i,\text{t}}$ is the conditional mean of the test data, and $\textbf{x}_{i(r),\text{t}}$ is the covariate vector of the $r$-th model under the test data.
	\end{enumerate}
	
	Algorithms 1 and 2, as well as the algorithm of the LEVSS approach with $T = 10$, is repeated $1, 000$ times, under each of the six cases. 
	
\begin{figure}[!thb]
		{\centering
			\includegraphics[width=1\textwidth]{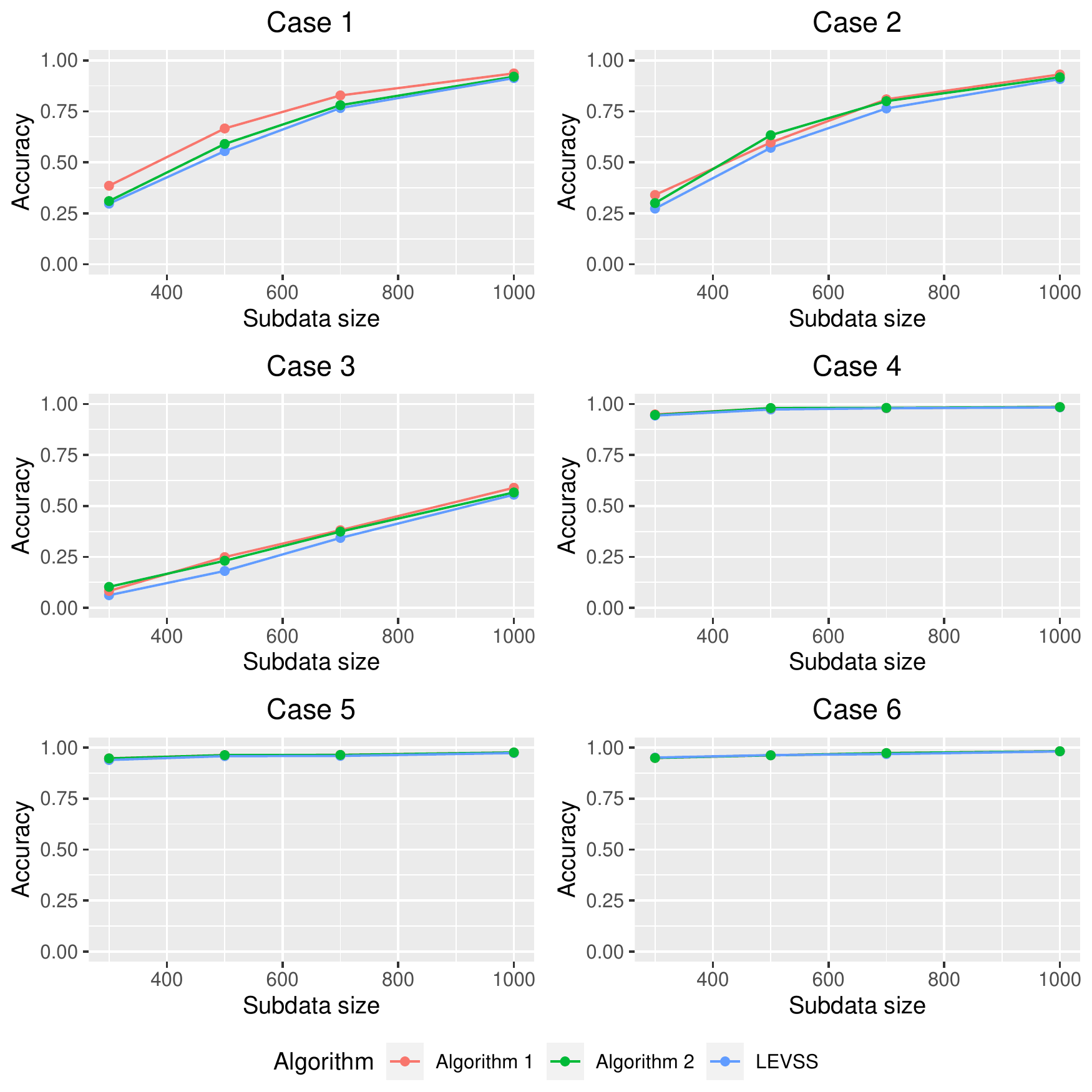}\par}
		\caption{Selection accuracies for different subdata sizes $k$ according to Algorithm 1, Algorithm 2 and LEVSS approach, for the six cases, based on all-subset regression via BIC.}
		\label{simMS}
	\end{figure}

	Figure \ref{simMS} presents the results on the accuracy. At first, the selection accuracies of Algorithms 1 and 2, as well as the LEVSS approach increase as $k$ increases. Also, Algorithms 1 and 2 outperforms the LEVSS approach in selecting the true model in Cases 1-3. All methods have a similar performance with high probabilities in selecting the true model in Cases 4-6. Moreover, Algorithm 1 performs better than Algorithm 2 in general. However, for some values of the subdata size in some cases, Algorithm 2 performs better than Algorithms 1. This fact shows the importance of Algorithms 1 and 2.
	
\begin{figure}[!thb]
		{\centering
			\includegraphics[width=1\textwidth]{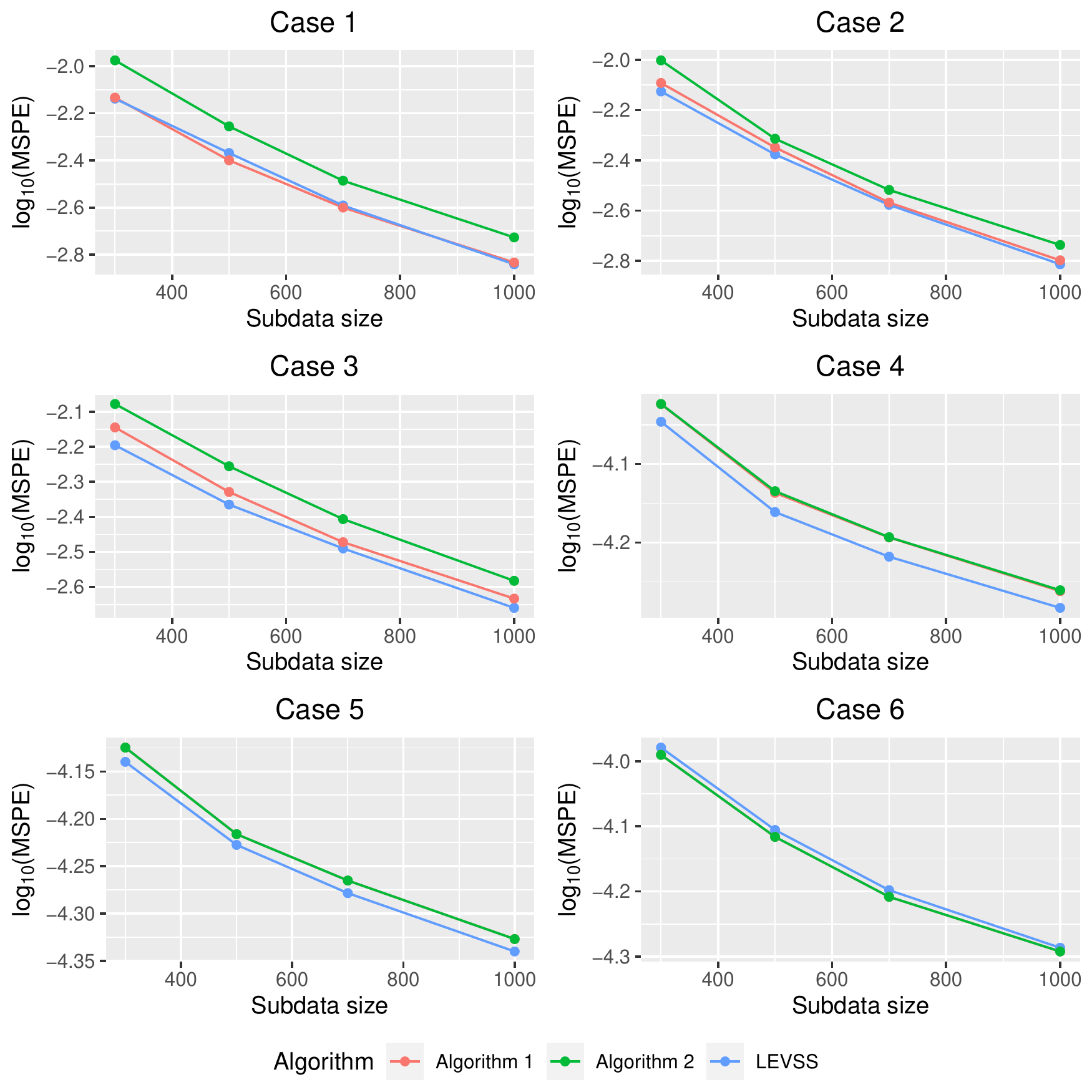}\par}
		\caption{MSPEs for different subdata sizes $k$ according to Algorithm 1, Algorithm 2 and LEVSS approach, for the six cases, based on all-subset regression via BIC. Logarithm with base $10$ is taken of MSPEs for better presentation of the figures.}
		\label{simP}
	\end{figure}

	Figure \ref{simP} presents the results on MSPE based on the selected model for $n_t=500$. The MSPE of Algorithms 1 and 2, as well as the LEVSS approach decrease as $k$ increases. For most of the cases examined, the LEVSS approach outperforms Algorithms 1 and 2 in minimizing the MSPE, even though Algorithms 1 and 2 have the advantage in selecting the true model compared with the LEVSS approach. A plausible reason for this is that the LEVSS approach is based on D-optimality criterion, and so provide more accurate estimates for the slope parameters. It is known in the current literature, that the selection of D-optimal subdata leads to more precise estimates of the unknown parameters (\cite{wang2019information}, \cite{chasiotis2023}, \cite{chasiotis2023levss}). However, a possibly higher selection accuracy of Algorithms 1 and 2 could lead to better results on MSPE.
	
	Now we focus on the computing time of Algorithms 1 and 2, and the algorithm of the LEVSS approach for different subdata sizes for Case 1, when the full data size is equal to $n=500,000$ and the number of covariates is equal to $p=7$. All computations are carried out on a PC with 3.6 GHz Intel 8-Core I7 processor and 16GB memory using the \texttt{R} programming language.
	
	In Table \ref{table1}, we present the mean computing times (in seconds) of Algorithms 1 and 2, and the algorithm of the LEVSS approach. 
	
	\begin{table}[!htb]
		\centering
		\setlength{\tabcolsep}{3pt}
		\begin{tabular}{ccccc}
			\toprule
			$k$ & $300$ & $500$ & $700$ & $1000$ \\ \midrule
			Algorithm 1 & 5.585 & 13.183 & 24.088 & 49.002 \\ \midrule
			Algorithm 2 & 6.730 & 14.963 & 26.491 & 50.528 \\ \midrule
			LEVSS & 1.114 & 1.124 & 1.149 & 1.159  \\ \bottomrule
		\end{tabular}
		\caption{The mean execution time (in seconds) of Algorithms 1 and 2, and the algorithm of the LEVSS approach for different subdata sizes $k=300, 500, 700$ and $1000$ for Case 1, when the full data size is equal to $n=500,000$ and the number of covariates is equal to $p=7$.}
		\label{table1}
	\end{table}
	
	The algorithm of the LEVSS approach is faster than Algorithms 1 and 2. This is expected due to Step 3 (main algorithm) of Algorithms 1 and 2. However, the time difference between our algorithms and the algorithm of the LEVSS approach should be considered in terms of the information gained. We remind that as we see in Figure \ref{simMS}, for example in Cases 1-3, Algorithms 1 and 2 outperforms the LEVSS approach in selecting the true model. Also, Algorithm 1 is slightly faster than Algorithm 2. Moreover, the computing times of Algorithms 1 and 2 can be reduced, if in the elimination step the design matrix \textbf{Q} consists of less than $2k$ data points.
	
	\section{Real data applications}
	\label{section_real_data}
	
	In this section, the performance of Algorithms 1 and 2 based on two real datasets, presenting the results of  the LEVSS approach as well, to make a comparison.
	
	\subsection{Diamonds}\label{diamonds}
	The first real data example is the \texttt{diamonds} dataset in the \texttt{ggplot2} package, which contains the prices and specifications for $53,940$ diamonds. To be more precise, following \cite{deldossi}, the seven features are:
	\begin{enumerate}
		\item The carat ($x_1$) that represents the weight of the diamond in the interval $[0.2, 5.01]$.
		\item The quality of the diamond cut ($x_2$), coded as $1$ if the quality is better than ``Very Good", and $0$ otherwise.
		\item The level of diamond color ($x_3$), coded as $1$ if the quality is better than ``level F", and $0$ otherwise.
		\item A measurement of the diamond clearness ($x_4$), taking the value $1$ if the quality is better than ``SI1", and $0$ otherwise.
		\item The total depth percentage ($x_5$).
		\item The width at the widest point ($x_6$).
		\item The volume of the diamond ($x_7$).
	\end{enumerate}
	
	Also, we do not consider $x_1$ in the analysis in order to avoid multicollinearity, since $x_1$ is highly correlated with $x_7$. Moreover, we include the quadratic effect of $x_7$ as $x_8$, and the response is the logarithm of the price ($\log_{10}$), in order to achieve a better fit of the data. Also, the dataset contains some outliers, which are excluded from the analysis, such as observation NO.24068, whose width is unusually large making the price too high.
	
	Assuming that any non-empty subset of the $p=7$ features can be a candidate set of active variables, there are $2^{7} - 1 = 127$ candidate models. Based on the full data, we consider as the true model the one with the smallest BIC value, that is:
	$$
	\hat{y} = 2.32469 + 0.01896x_2 - 0.08373x_3 + 0.12376x_4 - 0.00097x_6 + 0.01094x_7 -0.00002x_8
	$$
	
\begin{figure}[!thb]
		{\centering
			\includegraphics[width=1\textwidth]{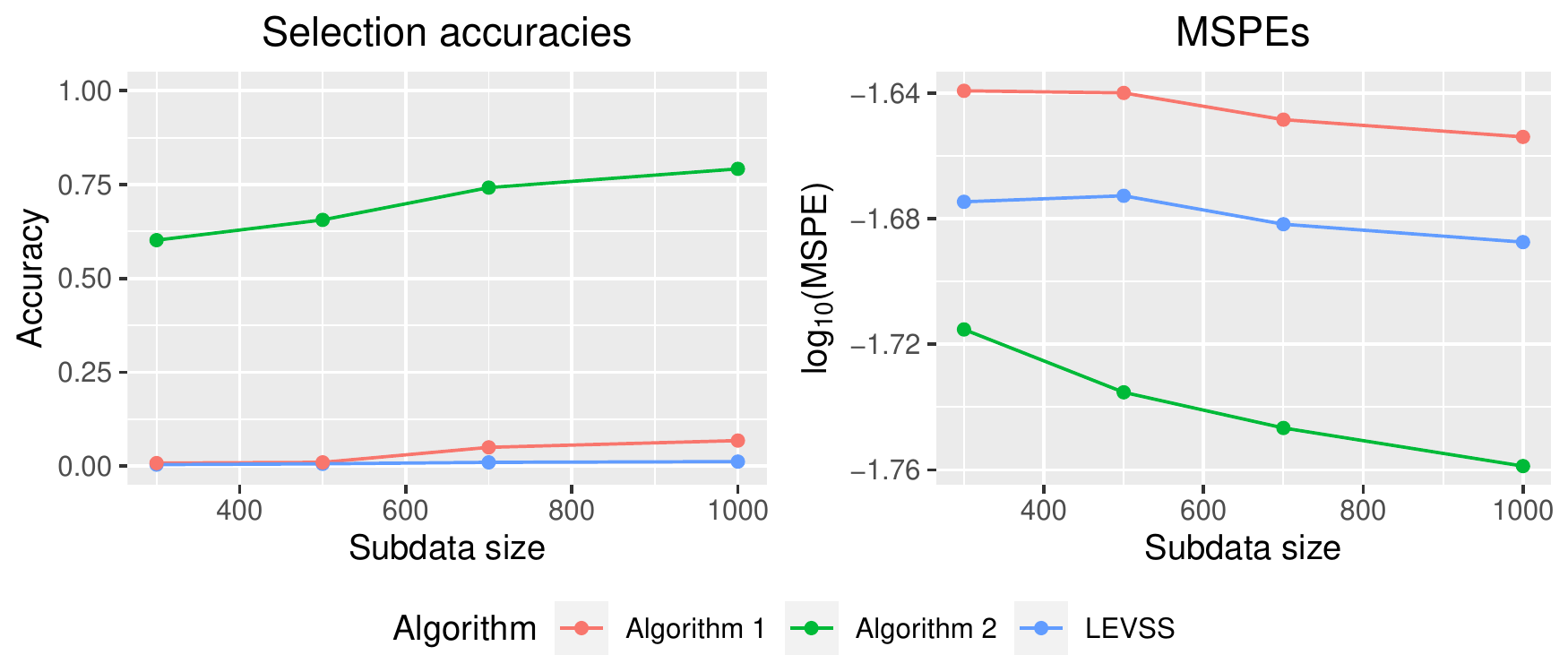}\par}
		\caption{Selection accuracies and MSPEs for different subdata sizes $k$ according to Algorithm 1, Algorithm 2 and LEVSS approach, for the diamonds dataset, based on all-subset regression via BIC. Logarithm with base $10$ is taken of MSPEs for better presentation of the figures.}
		\label{diamondsAS}
	\end{figure}

	Figure \ref{diamondsAS} presents the results on the accuracy and on MSPE based on the selected model according to all-subset regression. Algorithms 1 and 2 outperforms the LEVSS approach in selecting the true model. Also, Algorithm 2 has an exceptional performance both in selecting the true model and in minimizing the MSPE.
	
	Furthermore, we are interested in evaluating our approach based on Lasso regression, which is conducted through the \texttt{glmnet} package and the tuning parameters are selected through 10-fold cross-validation according to the \texttt{cv.glmnet()} function. Based on the full data, we consider as the true model the one according to Lasso regression, that is:
	\begin{align*}
		\hat{y} & = 2.30548 + 0.01832x_2 - 0.08263x_3 + 0.12214x_4 \\ &\quad + 0.00021x5 - 0.00069x_6 + 0.01082x_7 -0.00001x_8 
	\end{align*}
	
\begin{figure}[!thb]
		{\centering
			\includegraphics[width=1\textwidth]{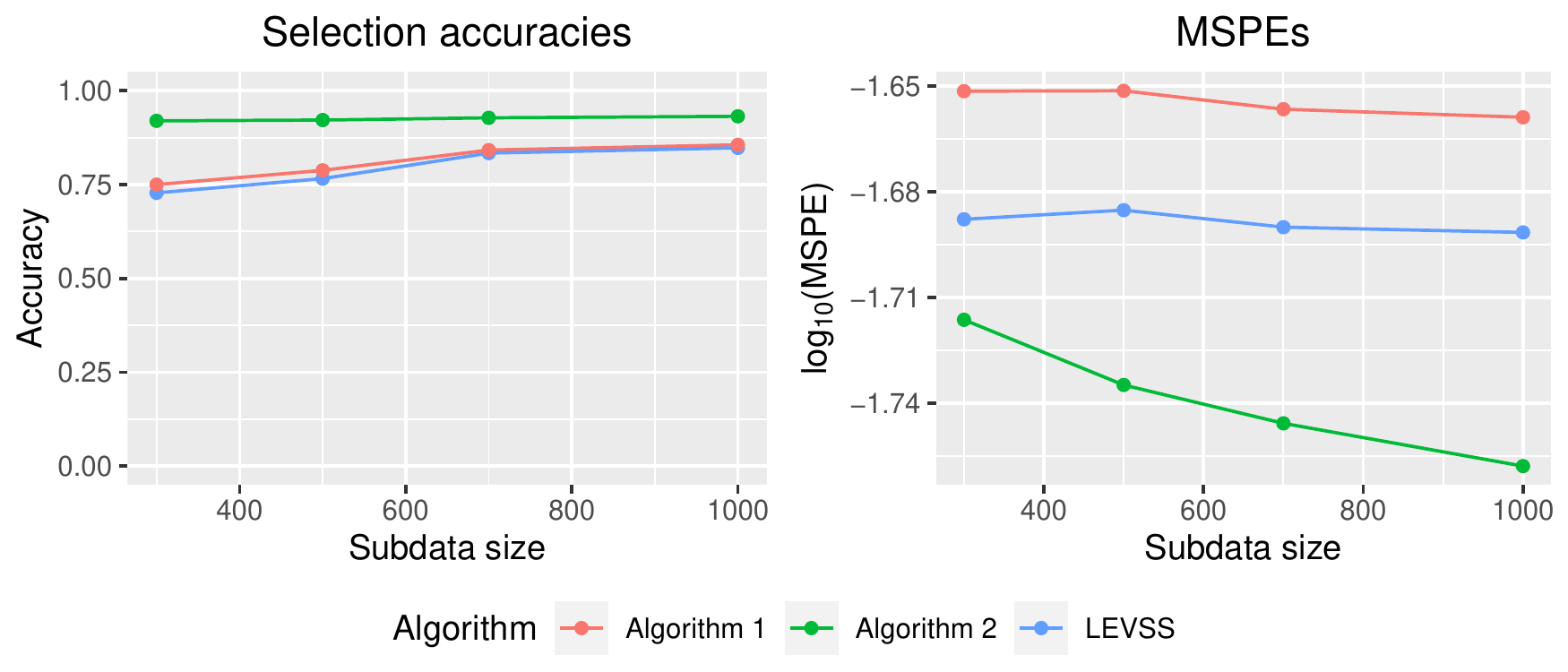}\par}
		\caption{Selection accuracies and MSPEs for different subdata sizes $k$ according to Algorithm 1, Algorithm 2 and LEVSS approach, for the diamonds dataset, based on Lasso regression. Logarithm with base $10$ is taken of MSPEs for better presentation of the figures.}
		\label{diamondsL}
	\end{figure}

	Figure \ref{diamondsL} presents the results on the accuracy and on MSPE based on the selected model according to Lasso regression. The results are very similar to the ones according to all-subset regression, that is Algorithms 1 and 2 outperforms the LEVSS approach in selecting the true model, and Algorithm 2 has a great performance both in selecting the true model and in minimizing the MSPE. Also, we need to mention that Algorithm 1 and the algorithm of the LEVSS approach, in selecting the true model, perform better in case of all-subset regression compared to the Lasso one.
	
	\subsection{Bias correction of numerical prediction model temperature forecast}
	The second real data example contains fourteen numerical weather prediction meteorological forecast data, two in-situ observations, and five geographical auxiliary variables over Seoul, South Korea in the summer from 2013 to 2017. To be more precise, this dataset is for the purpose of bias correction of next-day maximum and minimum air temperatures forecast of the LDAPS model operated by the Korea Meteorological Administration over Seoul, South Korea. Further information about the dataset can be found in ``UCI Machine Learning Repository" \cite{dua2019}.
	
	We select as response the next-day maximum air temperature. The full data contains $n=7, 590$ data points and $p=21$ features (station as well as date are excluded). Since the number of features is large to conduct an all-subset regression, we work based on Lasso regression, which is performed as in Section \ref{diamonds}. Based on the full data, we consider as the true model the one according to Lasso regression, and so none of the features are removed. To avoid the explanation of all variables, for more information see \cite{dua2019}, as well as for brevity, we do not provide the true model.
	
\begin{figure}[!thb]
		{\centering
			\includegraphics[width=1\textwidth]{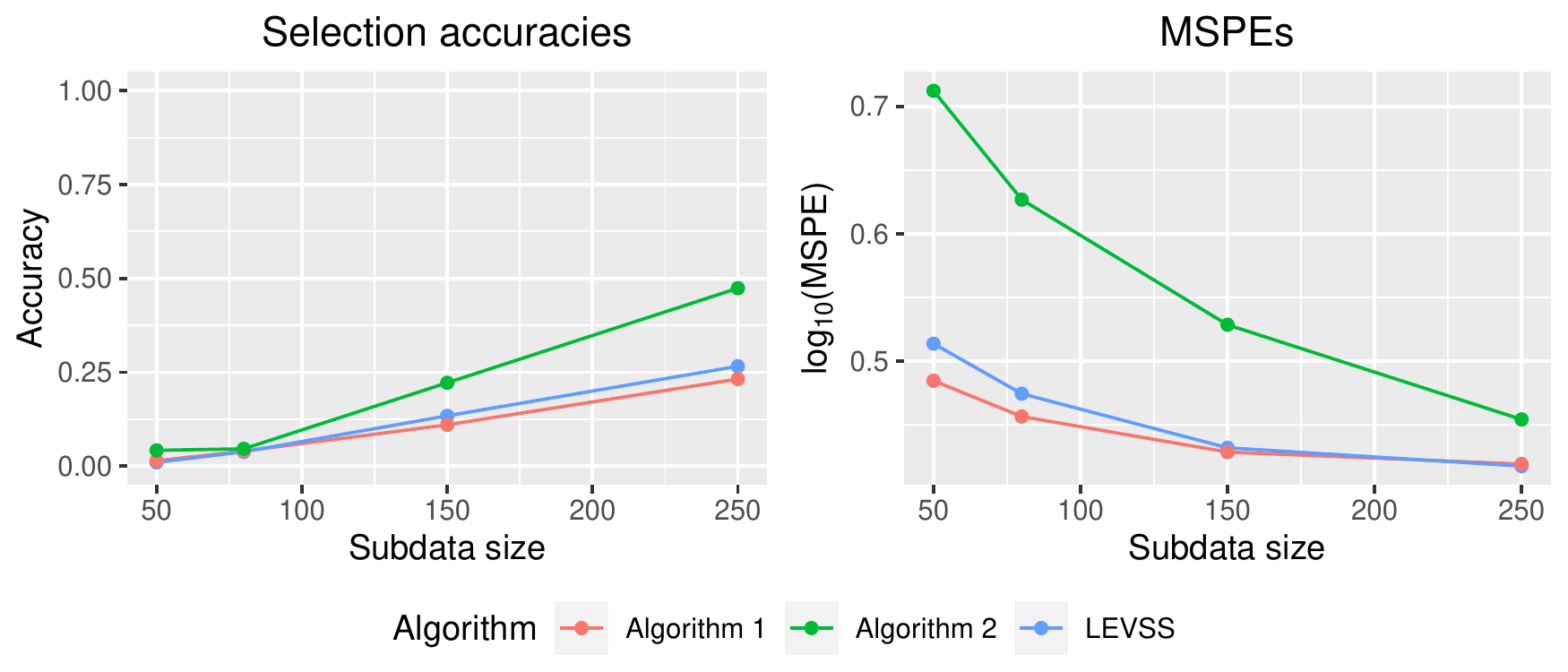}\par}
		\caption{Selection accuracies and MSPEs for different subdata sizes $k$ according to Algorithm 1, Algorithm 2 and LEVSS approach, for the bias correction dataset, based on Lasso regression. Logarithm with base $10$ is taken of MSPEs for better presentation of the figures.}
		\label{bias}
	\end{figure}

	Figure \ref{bias} presents the results on the accuracy and on MSPE based on the selected model according to Lasso regression. Algorithm 2 outperforms both Algorithm 1 and the LEVSS approach in selecting the true model, but Algorithm 1 outperforms both Algorithm 2 and the LEVSS approach in minimizing the MSPE.
	
	\section{Concluding remarks}
	\label{concluding_remarks}
	We presented two algorithms in order to select subdata from a big dataset based on the A-optimality criterion, so as to be able to conduct model selection. Our approach has been evaluated through simulations studies and two real data applications regarding the selection accuracy of the true model and the estimation efficiency, based on all-subset regression, Lasso regression and forward regression. 
	
	Our developed algorithms were compared with existing one of the LEVSS approach to show the kind of improvement gained. Our approach outperforms the LEVSS one in terms of selecting the true model. The results are also satisfactory regarding the estimation efficiency in the real data applications.
	
	We see that subdata selected based on the A-optimality criterion outperforms the ones based on the D-optimality criterion. However, we could look for some further improvement in the case of the estimation efficiency. Maybe a combination of both A-optimal and I-optimal designs could lead to the selection of subdata with even higher selection of the true model and much better estimation efficiency. I-optimality criterion focuses on prediction variance, and so seeks to minimize the average prediction variance over the design space. Due to this fact, \cite{deldossi} concluded that their proposed I-optimal algorithms should be applied in order to get accurate predictions on a specified prediction set.
	
	\section*{Appendix}
	As the number of variables ($p$) increase, the number of possible models increases exponentially, and so all-subset regression becomes impractical. In such cases, a common alternative is to use a forward selection approach, which starts with a model that does not include any variable, and iteratively adds one variable to the current ``best" model, based on the lowest BIC value. This process continues until no more variables can be added to improve the model. 
	
	We have chosen to adapt the forward regression method to illustrate our proposed approach. While backward elimination and stepwise regression can also be used, we focus solely on reporting the results obtained through forward regression due to their similar performance.
	
	\begin{figure}[!thb]
		{\centering
			\includegraphics[width=1\textwidth]{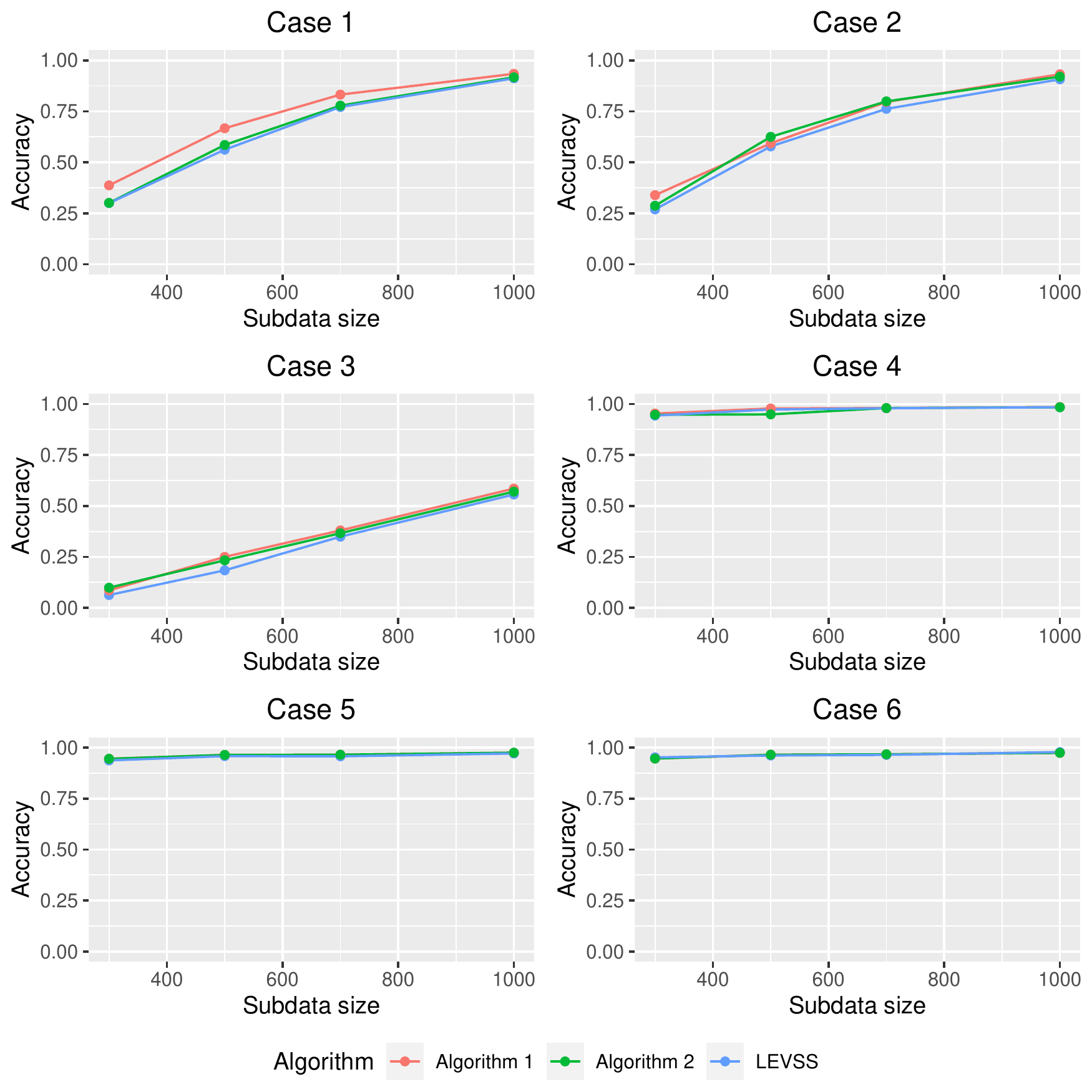}\par}
		\caption{Selection accuracies for different subdata sizes $k$ according to Algorithm 1, Algorithm 2 and LEVSS approach, for the six cases, based on forward regression via BIC.}
		\label{simMSf}
	\end{figure}
	
	\begin{figure}[!thb]
		{\centering
			\includegraphics[width=1\textwidth]{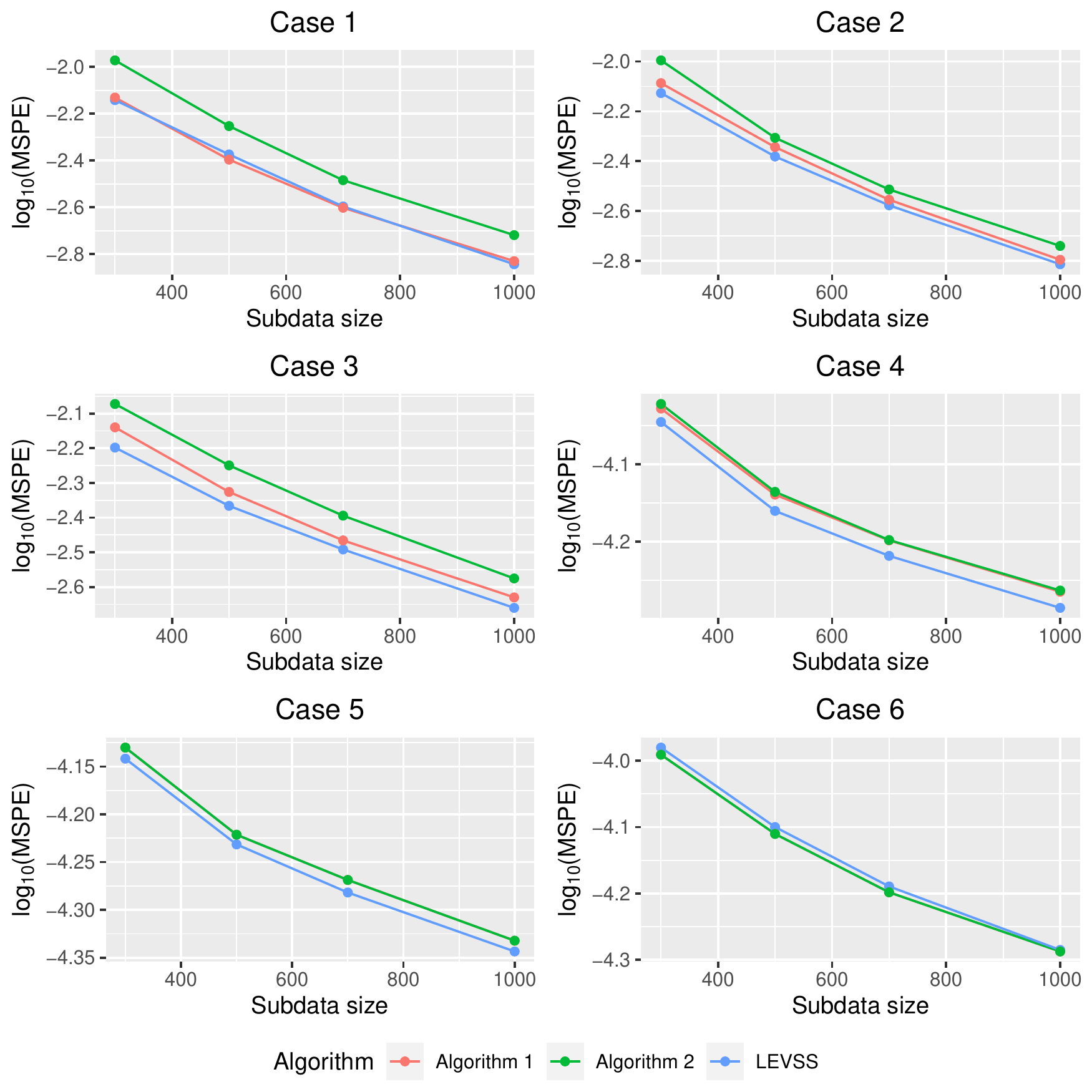}\par}
		\caption{MSPEs for different subdata sizes $k$ according to Algorithm 1, Algorithm 2 and LEVSS approach, for the six cases, based on forward regression via BIC. Logarithm with base $10$ is taken of MSPEs for better presentation of the figures.}
		\label{simPf}
	\end{figure}
	
	Figures \ref{simMSf} and \ref{simPf} present the results on the accuracy and on MSPE based on the selected model for $n_t=500$, respectively, based on forward regression via BIC. One can see that the results based on forward regression are very similar to the ones on all-subset regression.
	
	\bibliographystyle{apalike}
	\bibliography{main}
\end{document}